# Some applications of the Poisson summation formula

**M. L. Glasser**
Department of Physics, Clarkson University
Potsdam, NY 13699-5820 (USA)
laryg@clarkson.edu

**Nikos Bagis**
Department of Informatics, Aristotle University
Thessaloniki Greece
bagkis@hotmail.com

**Abstract**

In this note we derive two special forms of the Poisson's summation formula, for even and odd functions, which we apply to obtain representations for some Euler-type numbers $Q_n$, as well as to sum various series related to elliptic functions and elementary transcendental constants.

## 1. Introduction

One of the fundamental results of Fourier analysis is the Poisson Summation Formula, which can be written (see [Ch] pg. 43-47):

$$a\sum_{k=-\infty}^{\infty} f(ak) = \sum_{k=-\infty}^{\infty} f\wedge(bk) \qquad (1)$$

where $a > 0$, $a = 2\pi/a$ and

$$f\wedge(x) = \int_{-\infty}^{\infty} f(t)e^{-ixt}dt$$

is the Fourier Transform of $f$. Eq.(1) is valid under the relatively weak condition

$f(x) = O\left[\left(1+|x|^c\right)^{-1}\right]$, for $x \in \mathbb{R}$ and for some $c > 0$.

It is also known that if $f_c\wedge(\gamma) = \sqrt{\dfrac{2}{\pi}}\int_0^{\infty} f(t)\cos(t\gamma)dt$ is the Cosine Transform of a

function $f$ then someone can get the Poisson's Cosine Summation Formula (See [T] pg. 60-68):

$$\sqrt{a}\left(\frac{f(0)}{2} + \sum_{n=1}^{\infty} f(ka)\right) = \sqrt{b}\left(\frac{f_c\wedge(0)}{2} + \sum_{k=1}^{\infty} f_c\wedge(kb)\right) \qquad (2)$$

where $a > 0$, $b > 0$, $ab = 2\pi$.

## 2. Lemmas

**Lemma 1.**

Let $f$ is analytic in the upper half plane $\text{Im}(z) > 0$ and continuous in $\text{Im}(z) \geq 0$ and there exists $C, N>0$ and $0 \leq b < \pi$ such that

$$|f(z)| \leq C(1+|z|)^N e^{b|\text{Re}(z)|}, \text{ for every } z \text{ in } \text{Im}(z) \geq 0 \tag{3}$$

then the series $\sum_{k=0}^{\infty} (-1)^k f(i(k+1/2))$ is Abel summable and holds the following relation:

$$\int_{-\infty}^{\infty} \frac{f(t)}{\cosh(\pi t)} dt = 2 \lim_{r \to 1^-} \sum_{k=0}^{\infty} (-1)^k f(i(k+1/2)) r^k \tag{4}$$

If also $|f(ik)| \leq \dfrac{C'}{k+1}$, $k = 1, 2, 3, \ldots$ then the series converge and its sum is equal with the integral in Eq.(4).

**Lemma 2.**

Let $f$ is as in Lemma 1 then the series $\sum_{k=0}^{\infty} (-1)^k k f(ik)$ is Abel summable and holds the following relation:

$$\int_{-\infty}^{\infty} \frac{f(t)t}{\sinh(\pi t)} dt = 2 \lim_{r \to 1^-} \sum_{k=1}^{\infty} (-1)^{k-1} k f(ik) r^k \tag{5}$$

If also $|f(ik)| \leq \dfrac{C''}{k^2}$, $k = 1, 2, 3, \ldots$ then the series converge and its sum is equal with the integral in Eq.(5). (For a more general result of Lemma's 1, 2 see also [Bag]).

## 3. Theorems

**Theorem 1.**

Let $a > 0$, $f$ is even and as in Lemma 1, then

$$\sqrt{a}\left( \frac{f(0)}{2} + \sum_{k=1}^{\infty} \frac{f(ka)}{\cosh(\pi ka)} \right) = \sqrt{\frac{2b}{\pi}} \left( \frac{ce}{2} + \sum_{k=0}^{\infty} \frac{(-1)^k f(i(k+1/2))}{e^{b(k+1/2)} - 1} \right) \tag{6}$$

where $ab = 2\pi$ and $ce = \lim_{x \to 0^+} \left( \sum_{k=0}^{\infty} (-1)^k f(i(k+1/2)) e^{-(k+1/2)x} \right)$.

**Theorem 2.**

Let $a > 0$, $f$ is odd and as in Lemma 2, then

$$\sqrt{a}\left(\frac{f'(0)}{2\pi}+\sum_{k=1}^{\infty}\frac{f(ka)}{\sinh(\pi ka)}\right)=\sqrt{\frac{b2}{\pi}}\left(c_o/2+i\sum_{k=1}^{\infty}\frac{(-1)^k f(ki)}{e^{bk}-1}\right) \qquad (7)$$

when $ab=2\pi$ and $c_o=\lim_{x\to 0^+}\left(\sum_{k=1}^{\infty}(-1)^k if(ki)e^{-kx}\right)$.

## 4. Proof of Lemmas 1,2

**Proof of Lemma 1.**

Let $a>0$. We define the function

$$g(z)=e^{iza}f(z)\frac{1}{\cosh(\pi z)}$$

g is meromorphic in $\text{Im}(z)>0$ and has simple poles at the points $z=i(k+1/2)$ with

$\text{Res}(g;i(k+1/2))=e^{-a(k+1/2)}f(i(k+1/2))\frac{(-1)^k}{\pi i}$. Thus if $m>1$ natural, $R=m$ and $\gamma_R$ is the upper half circle with diameter [-R, R] we get:

$$\frac{1}{2\pi i}\int_{\gamma_{R_m}}g(z)dz=\sum_{k=0}^{m-1}\frac{(-1)^k}{\pi i}e^{-a(k+1/2)}f(i(k+1/2)) \qquad (8)$$

If m natural number we have $\left|\cosh\left(\pi \text{Re}^{i\theta}\right)\right|\geq ce^{\pi R|\cos(\theta)|}$, where c > 0 is a positive constant. From (3) we get for $0<a\leq \pi-b$:

$$\left|\int_0^{\pi}f(R\cdot e^{i\theta})\frac{e^{ia\text{Re}^{i\theta}}}{\cosh(\pi(\text{Re}^{i\theta}))}i\text{Re}^{i\theta}\,d\theta\right|\leq \left|\int_0^{\pi}(1+R)^N \text{Re}^{(b-\pi)R|\cos(\theta)|}e^{-aR|\cos(\theta)|}d\theta\right|\leq$$

$$\leq \int_0^{\pi}(1+R)^{N+1}e^{-aR}d\theta \to 0 \text{ as } R\to\infty \text{ }(R=m \text{ natural number}).$$

From (3) we have $\int_{-\infty}^{\infty}\left|f(t)\frac{1}{\cosh(\pi t)}\right|dt<\infty$ and $\sum_{k=0}^{\infty}e^{-ak}|f(i(k+1/2))|<\infty$

Thus we get $R=m\to\infty$ in Eq.(8) and we have:

$$\int_{-\infty}^{\infty}\frac{f(t)e^{ita}}{\cosh(\pi t)}dt=2\sum_{k=0}^{\infty}(-1)^k f(i(k+1/2))e^{-a(k+1/2)}, \text{ for every } a>0 \qquad (9)$$

The main result follows easily setting $a\to 0^+$ in Eq.(9).

**Proof of Lemma 2.**

As in Lemma 1.

# 5. Proof of Theorems

**Proof of Theorem 1.**

For $f$ even and as in Lemma 1 we have

$$\int_0^\infty \frac{f(t)\cos(t\gamma)}{\cosh(\pi t)}dt = \sum_{k=0}^\infty (-1)^k f(i(k+1/2))e^{-(k+1/2)\gamma} \qquad (10)$$

From Eq.(10) and the Poisson Cosine Formula Eq.(2):

$$\sqrt{a}\left(\frac{f(0)}{2} + \sum_{k=1}^\infty f(ka)\right) = \sqrt{b}\left(\frac{f_c{}^\wedge(0)}{2} + \sum_{k=1}^\infty f_c{}^\wedge(kb)\right)$$

or

$$\sqrt{a}\left(\frac{f(0)}{2} + \sum_{k=1}^\infty \frac{f(ka)}{\cosh(\pi ka)}\right) = \sqrt{b}\left(\frac{c e}{2} + \sum_{k=0}^\infty \frac{(-1)^k f(i(k+1/2))}{e^{b(k+1/2)} - 1}\right)$$

Where we have used that $\sum_{l=0}^\infty e^{-(k+1/2)lb} = \frac{1}{e^{b(k+1/2)} - 1}$.

**Proof of Theorem 2.**

From Lemma 2 we have $\int_{-\infty}^\infty \frac{f(t)te^{ita}}{\sinh(\pi t)}dt = 2\sum_{k=1}^\infty (-1)^{k-1} kf(ik)e^{-ka}$, $a > 0$

If $g$ is even then

$$\int_0^\infty \frac{g(t)t}{\sinh(\pi t)}\cos(ta)dt = \sum_{k=1}^\infty (-1)^{k-1} kg(ik)e^{-ka} \qquad (11)$$

From (1) and the Poisson cosine formula we have

$$\sqrt{a}\left(\frac{1}{2}\lim_{h\to 0}\left(\frac{g(h)h}{\sinh(\pi h)}\right) + \sum_{n=1}^\infty \frac{g(ka)ka}{\sinh(ka)}\right) = \sqrt{b}\left(\sqrt{\frac{2}{\pi}}c_e/2 - \sum_{m=1}^\infty 2\sqrt{\frac{2}{\pi}}\sum_{k=1}^\infty (-1)^{k-1} kg(ki)e^{-kmb}\right)$$

or

$$\sqrt{a}\left(\frac{1}{2}\lim_{h\to 0}\left(\frac{g(h)h}{\sinh(\pi h)}\right) + \sum_{n=1}^\infty \frac{g(ka)ka}{\sinh(ka)}\right) = \sqrt{\frac{b2}{\pi}}\left(c_e/2 - 2\sum_{k=1}^\infty \frac{(-1)^{k-1} kg(ki)}{e^{bk} - 1}\right)$$

or

$$\sqrt{a}\left(\frac{f(0)}{2\pi} + \sum_{n=1}^\infty \frac{g(ka)ka}{\sinh(ka)}\right) = \sqrt{\frac{b2}{\pi}}\left(c_e/2 - 2\sum_{k=1}^\infty \frac{(-1)^{k-1} kg(ki)}{e^{bk} - 1}\right)$$

The result follows replacing $f(x) = x\,g(x)$

# 6. Applications of Theorems 1,2

**1)** As a first application we take $f(t) = t^{2m+1}$ in Theorem 2. Since

$$c_o = (-1)^m \lim_{x \to 0^+} \sum_{k=1}^{\infty} (-1)^{k+1} k^{2m+1} e^{-kx} =$$

$$(-1)^{m+1} \lim_{x \to 0^+} \frac{d^{2m+1}}{dx^{2m+1}} \sum_{k=1}^{\infty} (-1)^{k+1} e^{-kx} = (-1)^{m+1} \lim_{x \to 0} \frac{d^{2m+1}}{dx^{2m+1}} \left( \frac{1}{e^x + 1} \right) =$$

$$(-1)^{m+1} \lim_{x \to 0} \frac{d^{2m+1}}{dx^{2m+1}} \left( \sum_{k=0}^{\infty} \frac{Q_k}{k!} x^k \right) =$$

$$\frac{(-1)^{m+1}}{2} Q_{2m+1} \qquad (12)$$

Theorem 2 therefore yields

$$\frac{a}{2} \left[ \frac{\delta_{m,0}}{2\pi} + a^{2m+1} \sum_{k=1}^{\infty} \frac{k^{2m+1}}{\sinh(\pi a k)} \right] =$$

$$\frac{(-1)^{m+1}}{4} Q_{2m+1} + (-1)^m \sum_{k=1}^{\infty} \frac{(-1)^{k+1} k^{2m+1}}{e^{2\pi k/a} - 1}. \qquad (13)$$

In particular, for $m = 0$ the hyperbolic series on the right hand side of (13) can be expressed in terms of elliptic functions [Z] which yields

$$\sum_{k=1}^{\infty} \frac{(-1)^{k+1} k}{e^{2\pi k/a} - 1} = \frac{1}{4}\left(\frac{1}{2} - \frac{a}{\pi}\right) + \frac{K}{\pi^2}[E - K] \qquad (14)$$

Where $K$ and $E$ are the complete elliptic integrals of the first and second kind of modulus k given by $K'/K = a$. When $a$ is the square root of a rational number, $k$ is one of the so-called singular moduli and the elliptic integrals can be expressed in terms of the Gamma function.

**2)** Next, we select $f(t) = t^{4n}$, $\mathbb{N} \ni n > 1$, for which, as above, we find $c_e - 1/2 Q_{4n+1}$. Then, taking a = 1, Theorem 2 gives

$$\sum_{k=1}^{\infty} \frac{k^{4n+1}(e^{k\pi} + (-1)^k)}{(e^{k\pi} - 1)(e^{k\pi} + 1)} = -1/4 Q_{4n+1} \qquad (15)$$

By using partial fractions and the identity ([Be2] pg. 262):

$$\sum_{k=1}^{\infty} \frac{k^{4n+1}}{e^{2\pi k} - 1} = \frac{B_{4n+2}}{8n+4}, \qquad (16)$$

where $B_n$, $n = 0,1,2,3,\ldots$ denotes the Bernoulli numbers, we find

$$\sum_{k=1}^{\infty} \frac{(2k-1)^{4n+1}}{e^{\pi(2k-1)} + 1} = -1/4 Q_{4n+1} - \frac{2^{4n-1}}{2n+1} B_{4n+2}. \qquad (17)$$

**3)** Other identities are:

**a)** Setting $f(t) = \dfrac{\sinh\left(\dfrac{t\pi}{6}\right)^6}{t^6}$ in Theorem 2, for $a = \pi$, $b = 2$ we get the result:

$$\frac{41\zeta(5)}{6912} = \frac{\pi^6}{93312} + 2\sum_{k=1}^{\infty}\frac{(-1)^k \sin^6\left(\dfrac{k\pi}{6}\right)}{k^5(e^{2k}-1)} + \frac{1}{\pi^4}\sum_{k=1}^{\infty}\frac{\sinh^6\left(\dfrac{k\pi^2}{6}\right)}{k^5 \sinh(\pi^2 k)}$$

Which is a formula for $\zeta(5)$.

**b)** $\dfrac{7\zeta(3)}{128} = \dfrac{\pi^3}{512} + \sum_{k=1}^{\infty}\dfrac{(-1)^k \sin^4\left(\dfrac{k\pi}{4}\right)}{k^3(e^{\pi k}-1)} + \dfrac{1}{8}\sum_{k=1}^{\infty}\dfrac{\sinh^4\left(\dfrac{k\pi}{2}\right)}{k^3 \sinh(2\pi k)}$

**c)** For $f(t) = \left(\dfrac{\cos(tc)-1}{t}\right)^3$, $a=1$, $b=2\pi$ we get:

$$8\sum_{k=1}^{\infty}\frac{(\cos(ck)-1)^3}{k^3 \sinh(k\pi)} + 16\sum_{k=1}^{\infty}\frac{(-1)^k(\cosh(ck)-1)^3}{k^3(e^{2k\pi}-1)} =$$

$$= -c^3 + c\pi^2 - 30Li_3(-e^{-c}) + 12Li_3(-e^{-2c}) - 2Li_3(-e^{-3c}) - 15\zeta(3) \quad : (18)$$

where $Li_v(x) = \sum_{k=1}^{\infty}\dfrac{x^k}{n^v}$, see [Be1] pg. 246).

Set now
$$f(c) := -30Li_3(-e^{-c}) + 12Li_3(-e^{-2c}) - 2Li_3(-e^{-3c}), \quad c \geq 0$$

Then we have the analytic continuation of $f$:

$$f(-c) = -2c^3 + 2c\pi^2 + f(c)$$

**d)** $\dfrac{1}{8}\sec^2\left(\tfrac{1}{2}\right)\tan\left(\tfrac{1}{2}\right) = 4\sum_{k=1}^{\infty}\dfrac{k^2 \sinh(2k)}{\sinh(2k\pi)} - \sum_{k=1}^{\infty}\dfrac{(-1)^k k^2 \sin(k)}{e^{\pi k}-1}$

**e)** $\dfrac{1}{16}(2-\cos(1))\sec^4\left(\dfrac{1}{2}\right) = 8\sum_{k=1}^{\infty}\dfrac{k^3 \sinh(2k)}{\sinh(2k\pi)} - \sum_{k=1}^{\infty}\dfrac{(-1)^k k^3 \cos(k)}{e^{\pi k}-1}$

**f)** $\tan\left(\dfrac{1}{2}\right) = \dfrac{2}{\pi} + 4\sum_{k=1}^{\infty}\dfrac{\sinh(2k)}{\sinh(2k\pi)} + 4\sum_{k=1}^{\infty}\dfrac{(-1)^k \sin(k)}{e^{\pi k}-1}$

**4)** If we set $f(t) = t^{4v+1} \cos(\pi t)$, $v = 1, 2, 3, \ldots$ and $a = 1$, $b = 2\pi$ Theorem 2 then

$$\sum_{k=1}^{\infty} \frac{(-1)^k k^{4v+1} \cosh(k\pi)}{e^{2k\pi} - 1} = -\frac{c_o}{2}$$

**i)** For $v = 1$ we get

$$\sum_{k=1}^{\infty} \frac{(-1)^k k^5 \cosh(k\pi)}{e^{2k\pi} - 1} = \frac{1}{64}(33 - 26\cosh(\pi) + \cosh(2\pi)) \sec h^6\left(\frac{\pi}{2}\right)$$

**ii)** For $v = 2$ we get

$$\sum_{k=1}^{\infty} \frac{(-1)^k k^9 \cosh(k\pi)}{e^{2k\pi} - 1} =$$

$$\frac{-e^\pi(1 - 502e^\pi + 14608e^{2\pi} - 88234e^{3\pi} + 156190e^{4\pi} - 88234e^{5\pi} + 14608e^{6\pi} - 502e^{7\pi} + e^{8\pi})}{(e^\pi + 1)^{10}}$$

**iii)** For $v = 3$ …etc. (All these sums are rational functions of $e^\pi$ with integer coefficients)

**5)** If we set $f(t) = \frac{\sin(\pi tv)}{t}$, $a = 1/v$, $b = 2\pi v$ in Theorem 1 then with $v > 0$, $v$-real, we get:

$$\operatorname{arccot}\left(e^{v\pi/2}\right) = 2\sum_{k=0}^{\infty} \frac{(-1)^k \sinh\left(v\pi(k + 1/2)\right)}{\left(e^{(2k+1)v\pi} - 1\right)(2k+1)}$$

**6)** If we set $f(t) = \left(\frac{\sin(\pi t/v)}{t}\right)^2$, $a = 1/v$, $b = 2\pi v$ in Theorem 1 then we get:

$$\tfrac{1}{4}e^{\pi/v}\varphi(-e^{2\pi/v}, 2, \tfrac{1}{2}) + \tfrac{1}{4}e^{-\pi/v}\varphi(-e^{-2\pi/v}, 2, 1/2) = 2G + \frac{\pi^2}{2v^3} -$$

$$-8\sum_{k=1}^{\infty} \frac{(-1)^k \sinh^2\left(\pi(\tfrac{k}{v} + \tfrac{1}{2v})\right)}{\left(e^{(2k+1)\pi v} - 1\right)(2k+1)^2} + v\sum_{k=1}^{\infty} \frac{\sin^2\left(\frac{k\pi}{v^2}\right)}{\cosh(k\pi/v)k^2}$$

Where $G$ is Catalan's constant: $G = \sum_{k=0}^{\infty} \frac{(-1)^k}{(2k+1)^2}$ and $\varphi(z, s, a) = \sum_{k=0}^{\infty} \frac{z^k}{(a+k)^s}$ is LerchPhi function (see [Be1]).